# Nanoscale creep mechanism of clay through MD modeling with hexagonal particles

Zhe Zhang, Xiaoyu Song*

*Engineering School of Sustainable Infrastructure and Environment,*
*University of Florida, Gainesville, Florida 32611*

**Abstract**

In this article, we investigate the creep mechanism of clay at the nanoscale. We conduct the molecular dynamics (MD) modeling of clay samples consisting of hexagonal particles under compression and shear. The MD simulations include oedometer creep, shear creep, direct shear tests, and stress relaxation. The numerical results show that the nanoscale creep mechanism of clay is related to particle rotation, translation, and stacking under different loading conditions. The clay sample under creep shows two types of particle arrangements, i.e., the shifted face-to-face configuration and the face-to-edge configuration. The orientation angle of clay particles is computed to track the rotation of individual particles due to creep. The fabric variation of the clay under creep is characterized by the dihedral angle between the basal particle plane and the x-y plane and the order parameter. It is found that the factors affecting the microstructure variation of the clay aggregate include stress levels, loading rates, and particle sizes. In the nanoscale shear creep test, the creep process comprises three stages, i.e., primary, secondary, and tertiary. The microstructure change during creep depends on the initial alignment of clay particles. The clay creep under a more significant stress level results in a more considerable order parameter and a more orientated clay structure.

**1. Introduction**

Creep of soft matters (such as clay) is an essential problem in engineering and science (e.g., soft matter science and geotechnical earthquake and hazards engineering) [1–9]. For instance, Alonso [9] described the nature of creeping landslide motion and its evolution towards a generalized failure involving a substantial acceleration of the moving mass. Several case histories of landslides have illustrated the occurrence of creeping motion before first-time failures and rapid sliding [8]. Instrumental measurements of displacement versus time have been used to study the fault creep [10]. Meanwhile, understanding the shear creep characteristics of fault rock and clay coatings is critical to the safe construction and long-term stability analysis of open-pit coal slopes and dam foundations [11, 12]. We note that most classical studies on clay creep have been focused on modeling creep at the continuum scale [13–19]. Few studies have been done on understanding the mechanism of clay creep at the nanoscale. Recent studies of fault creep have shown the occurrence of clay nanocoatings on the surface of fault rocks. This observation reflects the significant role of clay nanoparticles in slow fault creep [20]. In this article, as a new contribution, we conduct molecular dynamics (MD) modeling of clay creep with hexagonal particles under compression and shear loading conditions.

The creep of clay is a multiscale problem across space and time. For instance, the creep in clay is strongly dependent on the micro-structural constitution of clays. The shear loading could induce the translation and rotation of clay platelets. During the shear creep, the onset of slip units may consist of groups of parallel flaky particles [21]. Previous studies have concluded that the macroscopic physical and mechanical properties of clay are primarily dependent on its microstructural state at the scale of the particles [22–26]. For example, Xie et al. [25] investigated the creep behavior of loess-like soil using triaxial creep tests and SEM (scanning electron microscopy) image analysis. Zhao et al. [26] analyzed creep dilatancy and contracting phenomena using SEM and found that the creep dilatancy phenomenon is related to the expansion of micropores and micro-cracks within the overconsolidated clay specimen. Nanoindentation has also been applied to analyze the creep behavior of various materials such as clay shale and concrete [27–29]. However, few studies were devoted to understanding the creep mechanisms of clay particles at the atomistic scale. With the advance in high-performance computing, molecular dynamics becomes a viable tool to study clay's physical and mechanical properties, including creep at the nanoscale. We note that the molecular clay models in most MD studies are limited to perfect infinite clay platelets rather than a disordered system representing the actual clay assemblages (e.g., [30–35]). The macroscopic behavior of clay is controlled by complex interparticle physical and chemical interactions at the nanoscale. Rigid plate-wall models have been used to quantify the van der Waals force between two clay particles [36]. Underwood and Bourg [37] carried out all-atom simulations containing many discrete clay particles to understand clay's mechanical and physical behavior. Volkova et al. [38] investigated the microstructural configurations of finite-size Kaolinite particles with complex surface chemistry. However, few studies concern the nanoscale clay creep through MD modeling of clay assemblages. To close this knowledge gap, we present an MD modeling of clay creep using clay samples comprising clay assemblages. In what follows, we present a brief review of MD modeling of creep of soft materials, including clay.

---

*Corresponding author
 Email address:* xysong@ufl.edu (Xiaoyu Song)

MD modeling has been extensively employed to investigate the creep behavior of metals and polymers [39–41]. Creep behavior and microstructure evolution can be observed at a high strain rate in MD simulation, although the creep at the continuum scale typically occurs on time scales beyond the reach of molecular dynamics. For instance, in [42], the authors studied polyethylene's primary, secondary, and tertiary creep. However, such creep investigation through MD is still rare in clay minerals partially due to the complex molecular structure of clayey soils and the inherent in-homogeneity of clay matrix. The commonly used technique in MD simulation to achieve 1D compression is the Parrinello-Rahman method [43]. Several studies regarding creep employ this method to apply constant stress in the direction of interest while maintaining zero pressure in other directions [44–47]. We note that the disadvantage of such MD simulations is the uniform deformation of molecular structure by changing the size of the periodic boundary box. It is found that the applied stress to a polycrystalline material results in steady-state creep, i.e., a rate-dependent inelastic deformation characterized by a constant strain rate. Factors such as pressure and temperature are considered when investigating the underlying mechanism for steady-state creep, which may involve the diffusion of atoms through the lattice, the thermally activated glide, or interfacial processes [39, 48]. For polymers, MD simulations have shown creep mechanisms are the unkinking of chains and the chain alignment along the direction of force [41]. While the MD has been used to study the creep of metals and polymers, the MD modeling of clay is rare partially due to the complex molecular structure of clay and computational cost. In this study, we conduct MD modeling of clay creep under compression and shear using clay samples consisting of hexagonal clay particles.

This study aims to provide an atomistic scale understanding of the creep mechanism of clay under compression and shear loading conditions. We hypothesize that the nanoscale creep mechanism in clay aggregate is mainly attributed to particle rearrangements under various loading conditions, such as particle rotation and translation. The stress level and particle size may also affect the particle configuration under creep. To test this hypothesis, we conduct a series of MD modeling of clay samples consisting of hexagonal clay particles. The MD model features inhomogeneous clay aggregates. The hexagonal particle could characterize the edge properties of clay. We interpret the creep mechanism in terms of clay particle reorientation and reconfiguration. The MD modeling consists of odometer creep testing, shear creep testing, and stress relaxation testing. We also study the impact of loading rates on the micro-structure variation of clay through direct shear testing. The remainder of this article is presented as follows. Section 2 introduces the clay samples with hexagonal particles for the creep odometer and shear creep tests. Section 3 presents the numerical results of the MD modeling and analyzes the nanoscale creep mechanism of clay in terms of microstructure variation in creep. Section 4 is a summary of the research findings of this study.

## 2. MD models and simulations

In this work, we conduct four types of numerical tests: oedometer creep, shear creep, direct shear, and stress relaxation through MD simulations. The clay sample for each test is an assemblage of hexagonal pyrophyllite particles. The chemical formula of a pyrophyllite unit cell is $Al_2[Si_4O_{10}](OH)_2$. To accurately describe the edge structures of clay minerals, we use the hexagonal clay particle [49–51] instead of the rectangular clay particle in previous MD simulations [31, 32]. We construct a clay aggregate of hexagonal particles to ensure the microstructure is consistent with the actual clay stacking structure observed from a scanning electron microscope. Figure 1 shows a typical microstructure of stacking kaolinite platelets [52].

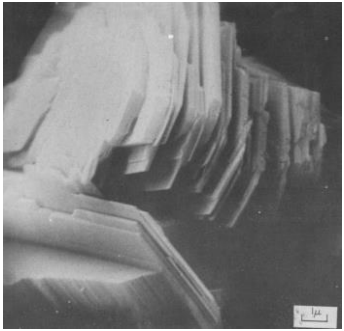

Figure 1: Stacking kaolinite platelets observed from scanning electron microscope [52].

Figure 2 shows the procedure for creating a hexagonal clay particle molecular model. We first build a one-layer rectangular clay particle consisting of 15 × 7 × 1 pyrophyllite unit cells. The dimensions are 77.4 Å, 62.762 Å, and 9.347 Å in the x, y, and z directions, respectively. To cut hexagons of various sizes out of the rectangle, we mark the boundaries of different hexagons sharing the same center in black. Different boundaries indicate different diameters of the hexagonal particles. In this work, we cut two hexagonal particles with different dimensions to investigate the particle size effect on clay creep behavior. The maximal diameter $D$ corresponds to the long diagonal distance of the hexagon. For the large particle, $D$ is 5.68 nm, and for the small particle, $D$ is 3.87 nm. For notation simplicity, we denote the large particle as a type-1 particle and the small particle as a type-2 particle.

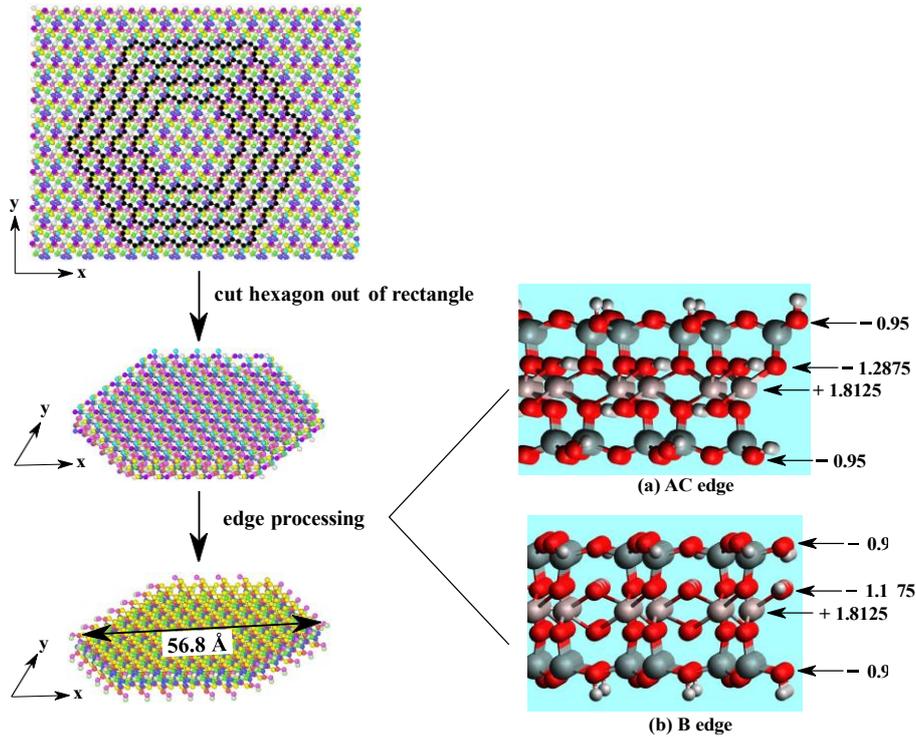

Figure 2: Schematic of the procedure for creating a hexagonal clay particle.

For each clay particle, six unstable edges, namely four [110] edges and two [010] edges (the so-called AC and B edges, respectively), are cleaved along crystallographic orientations to yield the most stable edges [49]. This is achieved by healing edges with -OH and -H groups to mimic near-neutral pH conditions [37, 50, 51]. The Lennard-Jones parameters for edge atoms are identical to their corresponding ClayFF parameters [53]. Upon healing the edges, we add hydrogen atoms to under-coordinated oxygen atoms. The partial charges of non-oxygen atoms along the clay edges are consistent with atomic charges defined by

ClayFF. However, the partial charge of edge oxygen atoms $q_O^{edge}$ is expressed as

$$q_O^{edge} = -2 + \sum_i \frac{q_i - q_i^C}{c_i}, \quad (1)$$

where $q_i$ is the valence of the $i$th neighboring atom of the edge oxygen, $q_i^C$ is the ClayFF assigned atomic charge of the $i$th neighboring atom, and $c_i$ is the coordination number of the $i$th neighboring atom.

MD simulations are conducted on a massively parallel molecular simulator [54]. Atomic interactions are described through the ClayFF force field [53] with a cutoff radius $r_c$ that equals 1 nm. The total potential energy consists of four terms, van der Waals energy described by Lennard-Jones 12-6 potential, Coulomb electrostatic potentials calculated with partial charges, bond stretch energy, and angle bend energy in harmonic form. We treat the nanoparticle as a rigid body assuming the incompressibility of the soil skeleton, which also improves computational efficiency. Ignoring the deformation in individual particles, we focus on the changes in particle orientation and microstructure of clay aggregate during creep. Since we employ the rigid clay particle, only nonbonded interactions are active throughout the simulations. The van der Waals energy term is written by the Lennard-Jones 12-6 function as

$$E_{vdw} = \sum_{i \neq j} D_{ij} \left[ \left(\frac{R_{ij}}{r_{ij}}\right)^{12} - 2 \left(\frac{R_{ij}}{r_{ij}}\right)^6 \right], \quad (2)$$

where $r_{ij}$ is the distance between atoms $i$ and $j$, and $D_{ij}$ and $R_{ij}$ are respectively the depth of the potential well and the distance at which the interatomic potential energy vanishes. The Coulombic energy term is written as

$$E_{coul} = \frac{1}{4\pi\varepsilon_0} \sum_{i \neq j} \frac{q_i q_j}{r_{ij}}, \quad (3)$$

where $q_i$ and $q_j$ are atomic charges of atoms $i$ and $j$, and $\varepsilon_0$ is the vacuum permittivity. Note that edge atoms have different partial charges than non-edge atoms. The particle-particle particle-mesh method is used to compute the long-range Coulombic interactions. Periodic boundary conditions are applied in all directions. The Nosé-Hoover thermostat [55, 56] is used to control temperature, and the Parrinello-Rahman barostat [43] is used to control pressure in the molecular system. The coordinates, velocities, and orientations of atoms in the same rigid particle are updated synchronously so that the rigid body moves and rotates as a single entity. This assumption allows us to disable the interaction between two atoms in the same clay platelet. Next, we briefly introduce the clay model setups for the numerical creep tests in this study.

Table 1 summarizes all clay samples reported in this study. Here the article size refers to the maximal diameter $D$, i.e., the maximum diagonal distance of the hexagon.

Table 1: Summary of test samples. Note: Sample 3 consists of 289 large particles and 289 small particles.

| Sample | Particle size (nm) | Number of particles | Sample dimensions (nm$^3$) | Testing type |
|---|---|---|---|---|
| 1 | 5.68 | 427 | 35 × 35 × 35 | Oedometer creep, Shear creep |
| 2 | 3.87 | 889 | 35 × 35 × 35 | Oedometer creep |
| 3 | 5.68 and 3.87 | 578 | 35 × 35 × 35 | Oedometer creep |
| 4 | 5.68 | 523 | 50 × 15 × 70 | Direct shear test, Shear stress relaxation test |
| 5 | 3.87 | 1088 | 50 × 15 × 70 | Direct shear |

*2.1. Oedometer creep test*

The clay aggregate for the oedometer test consists of 427 type-1 particles. These particles are randomly packed in a periodic cubic simulation box with a side length of 35 nm. The initial packing density is 0.6 $g/cm^3$. Figure 3 (a) shows the boundary conditions of the clay sample in the oedometer creep test. The MD simulation of the clay creep test follows the procedure and setup of a laboratory oedometer test [57, 58]. We first obtain an equilibrated clay sample in the canonical (NVT) ensemble (i.e., constant number of particles, constant system volume, and constant temperature). Then a constant vertical stress equal to 100 MPa is applied to the sample in the z-direction. This guarantees that pressure control acts only along the vertical direction. In the 1-D creep test, only the height of the clay sample can change, and the sample is radially constrained. Since we assume each particle is a rigid body, the volumetric change during the oedometer test is purely attributed to changes in pore space without the solid particle volume change. All simulations are run on a supercomputer with 128 CPU cores. The wall clock time of the equilibration run is approximately 100 hours. The wall clock time for obtaining a steady-state creep is about 75 hours for the equilibrated sample.

*2.2. Shear creep test*

The same clay sample as the oedometer creep test is adopted for the shear creep test. In the shear creep test, the clay sample is first equilibrated in an isothermal-isobaric (NPT) ensemble at 298 K and 1 atm. Figure 3 (b) shows the boundary conditions of the clay sample in the shear creep test. A constant normal pressure $P_z = 10$ atm is prescribed on the clay aggregate. When the steady state has been obtained, and no further volume changes in the clay sample, the normal pressure holds, and a constant shear force is prescribed on the top and bottom boundary layers as shown in Figure 3 (b).

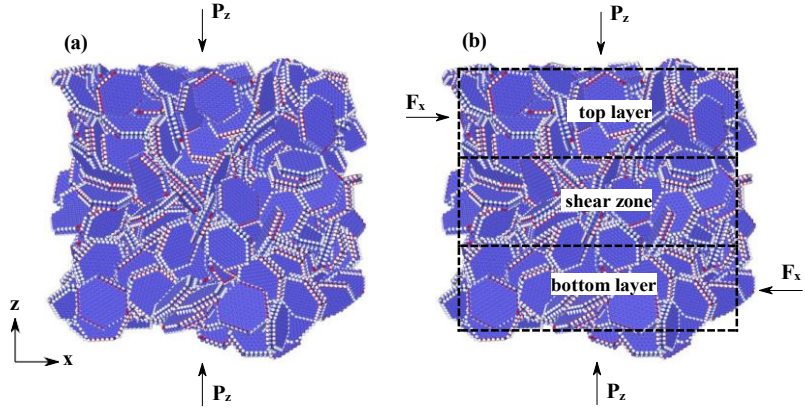

Figure 3: Boundary conditions of the clay samples in (a) the oedometer creep test and (b) the shear creep test.

*2.3. Direct shear and relaxation test*

For the direct shear test, the dimensions of the clay sample are 50 nm, 15 nm, and 70 nm in the x-, y-, and z-directions, respectively. For the base simulation, 523 type-1 particles are packed with an initial density of 0.6 $g/cm^3$. Figure 4 shows the setup of the direct shear test. An equilibrated configuration is first obtained in the NVT ensemble under 298 K. The clay aggregate is then equilibrated in the NPT ensemble under the constant vertical pressure of 10 atm. Then, the top layer is loaded horizontally at a constant velocity $v_x$. The bottom boundary layer is fixed during the simulation. The particles in the shear zone are free to translate and/or rotate. Four shear rates ranging from 0.001 Å /fs to 0.004 Å /fs are adopted to investigate the shear rate effect on the microstructure of the shear zone and the shear strength of the clay sample. In the following section, we present the numerical results of the MD simulations of the clay and analyze the nanoscale creep mechanism under odometer and shear conditions.

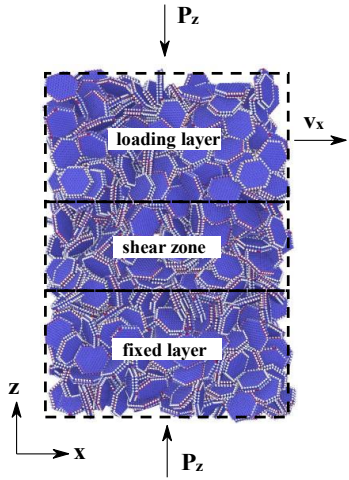

Figure 4: Boundary conditions of the clay sample in the direct shear test.

## 3. Numerical results of MD creep testing

This section presents the numerical results of MD creep tests, including oedometer creep, shear creep, and shear stress relaxation. We investigate the nanoscale creep mechanism of clay by analyzing the microstructure change during the creep testing. First, we present the results of oedometer creep testing, which includes the impact of stress level and particle size. Second, we discuss the results of shear creep testing. Third, we present direct shear testing to study the effects of loading rates on the microstructure change under direct shear. In the fourth part, we analyze the shear stress relaxation testing results.

### 3.1. Oedometer creep test

We first present the results of a base simulation of the oedometer creep test. Then we present the results of the impacts of stress level and particle sizes on the oedometer creep. Figure 5 shows the variation of the axial strain of the clay sample in the oedometer test under 100 MPa. The strain rate first increases dramatically upon loading and then gradually increases with loading time. The ultra-high strain rate is up to 17.9 $ns^{-1}$ in the initial linear section, followed by the steady-state creep. Next, we present the variation of microstructures of the clay aggregate in the oedometer creep testing. Figure 6 plots the cross-section profiles of the clay aggregate at six loading times in the oedometer test under 100 MPa. Initially, particles are loosely packed with random orientations. From (a) to (b), one observes a significant closure of micropores between particles. This structural change explains the initial high strain rate in Figure 5. The particle rearrangement occurs progressively after the distance between particles is smaller than the cutoff distance of Lennard-Jones interactions. It is found that particles form a stacking structure during the oedometer creep. The preferred orientation direction is horizontal, which agrees with the vertical loading direction.

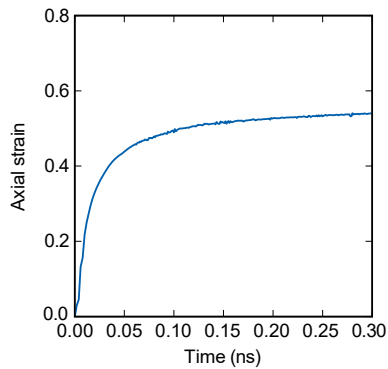

Figure 5: Variation of the axial strain with time in the oedometer test under 100 MPa.

We interpret the nanoscale creep mechanism by characterizing the time variation of local configurations of selected neighboring particles. In general, we find two major particle configurations, i.e., the shifted face-to-face configuration and the face-to-edge or angular configuration. The changes in the microstructure of the clay aggregate under the constant vertical stress can be divided into three stages: (a) The closure of pore space between particles, (b) the particle rearrangement due to pair interactions that form face-to-edge or face-to-edge particle configurations, and (c) the progressive slippage of interfaces between particles. Figure 7 presents the formation of the shifted face-to-face configuration in the oedometer creep test under 100 MPa. Upon loading, the pore volume between particles decreases rapidly, and particles get closer to each other. Once the center-to-center distance between particles is smaller than the interaction cutoff distance, inter-particle attraction plays a crucial role in particle stacking. As a result, neighboring particles gradually form a shifted face-to-face configuration. As the creep continues, we observe the progressive slippage of the particle interface with a larger interfacial area. Therefore, we may conclude that the optimal equilibrium configuration between two particles is a parallel face-to-face configuration. A shifted face-to-face configuration is observed due to the initial disordered particle orientations.

The other type of particle configuration observed is the face-to-edge configuration. Figure 8 presents the formation of a face-to-edge configuration. In the oedometer creep test, the initially vertically aligned particles tend to keep the initial alignment and exhibit the face-to-face arrangement with other vertically aligned particles. When interacting with horizontally aligned particles at the bottom, these particles form a local "locking zone," and the angular configuration remains stable during the creep testing. Comparing the two types of configurations indicates that the particle arrangement largely depends on the initial alignment of particles. Clay particle configurations from our MD simulations are similar to those shown in Figure 1. We note that Diamond [22] found both the parallel orientation of particles (i.e., face-to-face configuration) and particles tilted to each other (i.e., a face-to-edge configuration).

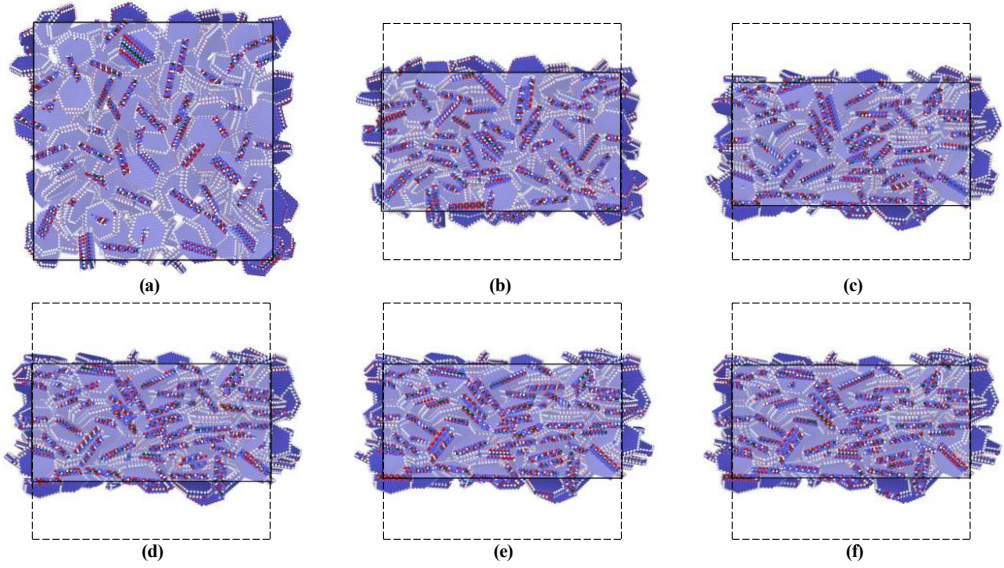

Figure 6: Configurations of the clay aggregate at six loading times: (a) t = 0 ns, (b) t = 0.04 ns, (c) t = 0.08 ns, (d) t = 0.12 ns, (e) t = 0.16 ns, (f) t = 0.20 ns in the oedometer creep test under 100 MPa.

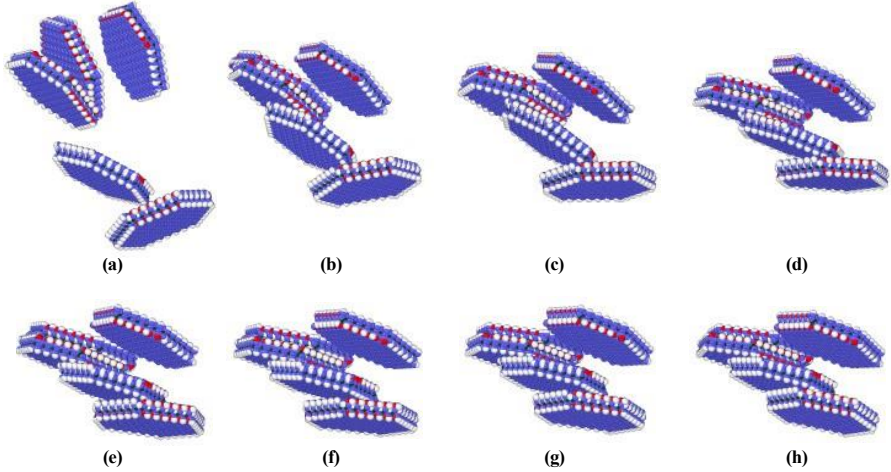

Figure 7: Snapshots of selected particles in the creep test under 100 MPa at eight loading times: (a) t = 0 ns, (b) t = 0.04 ns, (c) t = 0.08 ns, (d) t = 0.12 ns, (e) t = 0.16 ns, (f) t = 0.20 ns, (g) t = 0.24 ns, and (h) t = 0.28 ns that show the formation of the shifted face-to-face particle configuration.

The void volume is the difference between the specimen volume and the total clay particle volume. Figure 9 plots the time variations of porosity and density of the clay aggregate in the oedometer creep test under 100 MPa. The porosity at the loading time of 0.3 ns decreases by 25.3% of the initial porosity of 0.82. Due to creep, the sample density increases from 0.6 g/cm$^3$ to 1.3 g/cm$^3$. Figure 10 shows a surface representation of clay aggregate constructed using the alpha shape method [59]. The alpha shapes of the clay aggregate geometrically describe the outer boundary and connections of the inner pores of the clay sample. The surface area decreases from 2.34×10$^6$ Å$^2$ to 2.24×10$^6$ Å$^2$ at the loading time 0.03 ns. We note that the decrease in the surface area of clay aggregate is due to the formation of shifted face-to-face particle configurations during creep. In what follows, we present the results of the effects of stress levels and particle sizes on the oedometer creep.

### 3.1.1. Effect of stress levels

It is known that creep rate depends on stress [60]. In this part, we investigate the effect of stress on the nanoscale creep mechanism of clay by comparing the results of four oedometer creep tests under 10 MPa,

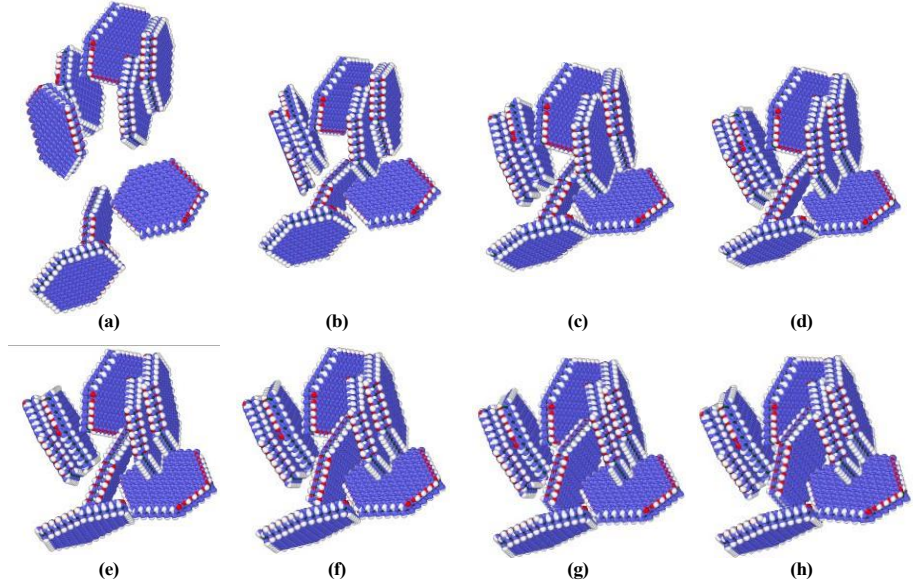

Figure 8: Snapshots of selected particles in the oedometer creep test under 100 MPa at eight loading times: (a) t = 0 ns, (b) t = 0.04 ns, (c) t = 0.08 ns, (d) t = 0.12 ns, (e) t = 0.16 ns, (f) t = 0.20 ns, (g) t = 0.24 ns, (h) t = 0.28 ns that show the formation of the face-to-edge particle configuration.

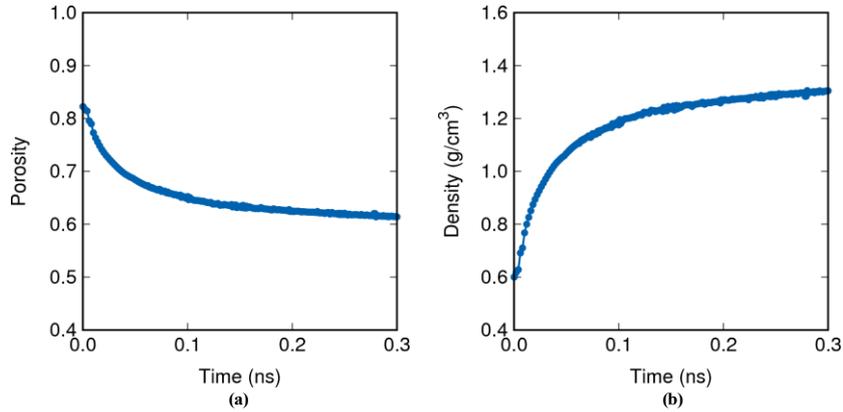

Figure 9: Variations of (a) porosity and (b) density of the clay sample in the oedometer creep test under 100 MPa.

100 MPa, 300 MPa, and 2000 MPa, respectively. Figure 11 shows the configurations of clay aggregates in the four creep tests at the same loading time of 0.3 ns. Figure 12 shows the time-averaged pressure $P_z$ in the clay samples under the four stress levels. After the initial oscillation, $P_z$ becomes constant during the remaining oedometer creep test. Figure 13 compares the variation of axial strains in clay samples under four stress levels. Higher vertical stress generates a larger strain rate at the beginning and accelerates the onset of the creeping stage. The creep stage corresponds to the linear section of the strain-time curve with an almost constant slope [39, 61]. It is assumed that the creep in clay occurs after 0.18 ns. Based on our MD simulations, the steady-state creep rate is about 0.145 ns$^{-1}$, 0.130 ns$^{-1}$, 0.111 ns$^{-1}$, and 0.048 ns$^{-1}$ for the stresses 10 MPa, 100 MPa, 300 MPa, and 2000 MPa, respectively. As time continues, the creep strain increases gradually, and a higher stress level leads to a larger creep deformation. For example, the axial strains at the same loading time of 0.3 ns are 0.52, 0.54, 0.57, and 0.59 under 10 MPa, 100 MPa, 300 MPa, and 2000 MPa, respectively. The time needed to reach 90% of the creep strain at t = 0.3 ns are computed and compared to show the stress effect on clay creep during the oedometer test. For the test under 10 MPa, it takes 0.114 ns to get 90% creep strain. Under the higher stress levels, the times are 0.092 ns, 0.075 ns, and 0.044 ns for 100 MPa, 300 MPa, and 2000 MPa, respectively.

We quantify the degree of particle orientations in the clay aggregate through the order parameter $S$. The order parameter denotes the number of orientation correlations between particles [62]. We first compute

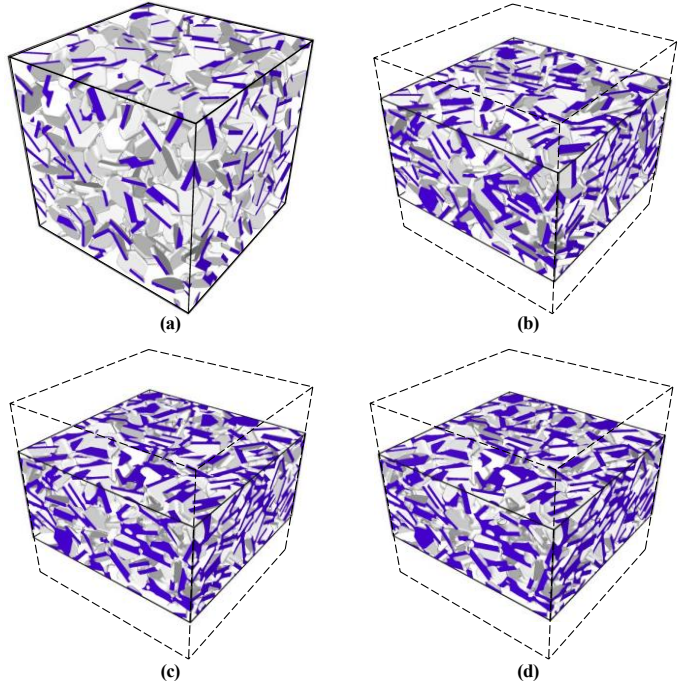

Figure 10: Alpha shape of the clay aggregate in the oedometer creep test under 100 MPa at four loading times: (a) t = 0 ns, (b) t = 0.01 ns, (c) t = 0.02 ns, and (d) t = 0.03 ns.

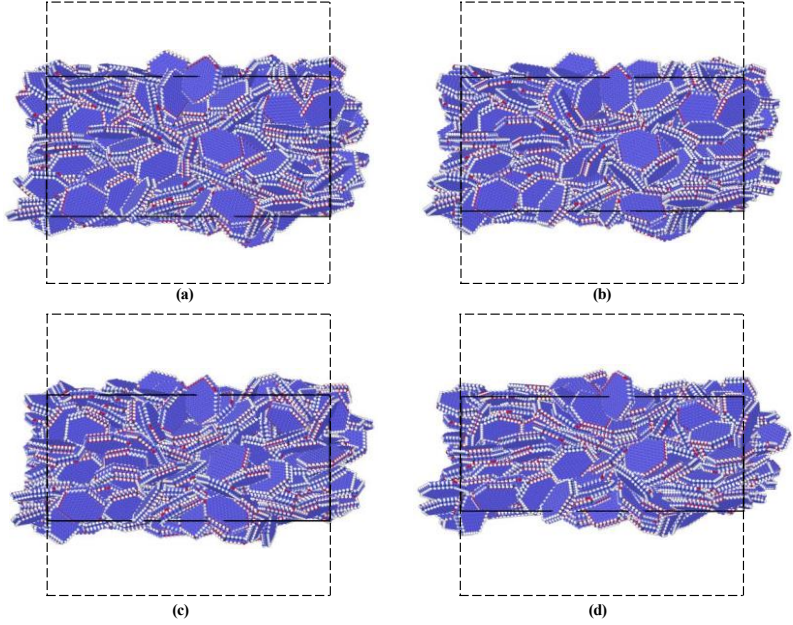

Figure 11: Configurations of the clay samples in the oedometer creep test under (a) 10 MPa, (b) 100 MPa, (c) 300 MPa, and (d) 2000 MPa at the same loading time 0.3 ns.

the director tensor of the clay system $Q_{ij}$ [37] that reads

$$Q_{ij} = \frac{1}{N}\sum_{k=1}^{N}\left(\frac{3}{2}\hat{n}_{ki}\hat{n}_{kj} - \frac{1}{2}\delta_{ij}\right) \quad \text{and} \quad i, j = 1, 2, 3, \qquad (4)$$

where $N$ is the number of particles in the system, $\hat{n}_k$ is the normal unit vector of the basal surface of the $k$th particle, and $\delta_{ij}$ is the Kronecker delta function. The scalar order parameter $S$ is the largest absolute eigenvalue of the director tensor $Q_{ij}$. For a completely random particle orientation, $S$ is 0, and $S = 1$ for a perfectly aligned clay aggregate. Figure 14 shows the variation of the order parameter $S$ of the clay sample under four stress levels. The order parameter first increases dramatically, followed by a decreasing changing rate. The steady-state linear change in the order parameter corresponds to the secondary creep where the shifted face-to-face configurations have formed. The difference in the order parameter $S$ during creep under different stress levels indicates that higher vertical stress produces a more significant amount of shifted face-to-face configurations that may accelerate the onset of creep. For example, the order parameter of the clay aggregate is up to 0.52 under 2000 MPa, while $S$ is only 0.42 for the clay sample under 100 MPa. At t = 0.3 ns, the clay samples under 300 MPa and 2000 MPa give almost the identical order parameter whose value is far smaller than that of the perfectly aligned configuration (i.e., $S = 1$). This may imply a unique upper bound of the order parameter for each clay aggregate regardless of the applied vertical stress. In other words, the initial clay particle orientation could determine the equilibrium configuration of the clay aggregate during creep.

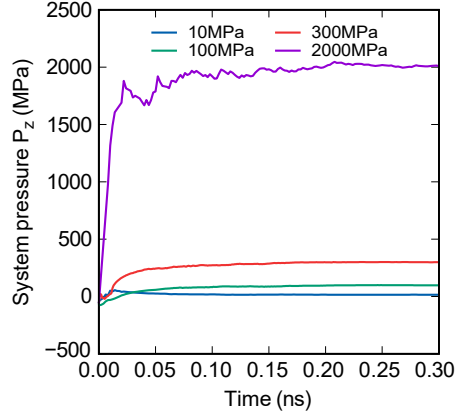

Figure 12: Time averaged pressure $P_z$ in the clay samples under different stress levels.

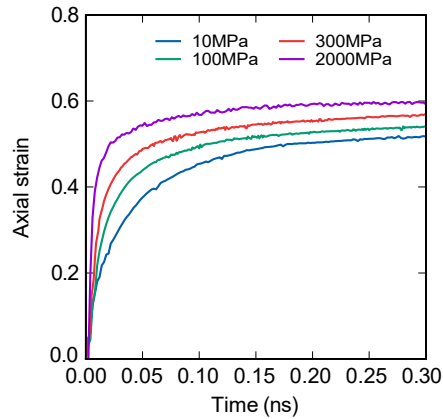

Figure 13: Variation of the axial strain in clay aggregates in the oedometer creep tests under the four stress levels.

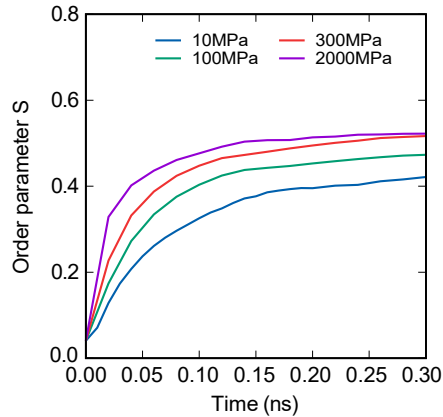

Figure 14: Variation of the order parameter of clay aggregate in the oedometer creep test under four stress levels.

Next, we present the orientations of individual particles to show the variation of particle configuration during the creep under different stress levels. Figure 15 plots the histogram of particle orientation angles under the four stress levels. The angle is defined as the angle between the clay basal surface and the z-axis. For example, if the clay particle is horizontally aligned, its orientation angle is 90°. The average particle orientation angles for 10 MPa, 100 MPa, 300 MPa, and 2000 MPa are 52.5°, 54.9°, 56.9°, and 57.1°, respectively. The larger orientation angle indicates more of the shifted face-face particle configurations. Figure 16 plots the variation of the axial strain with the logarithmic time during the oedometer creep test.

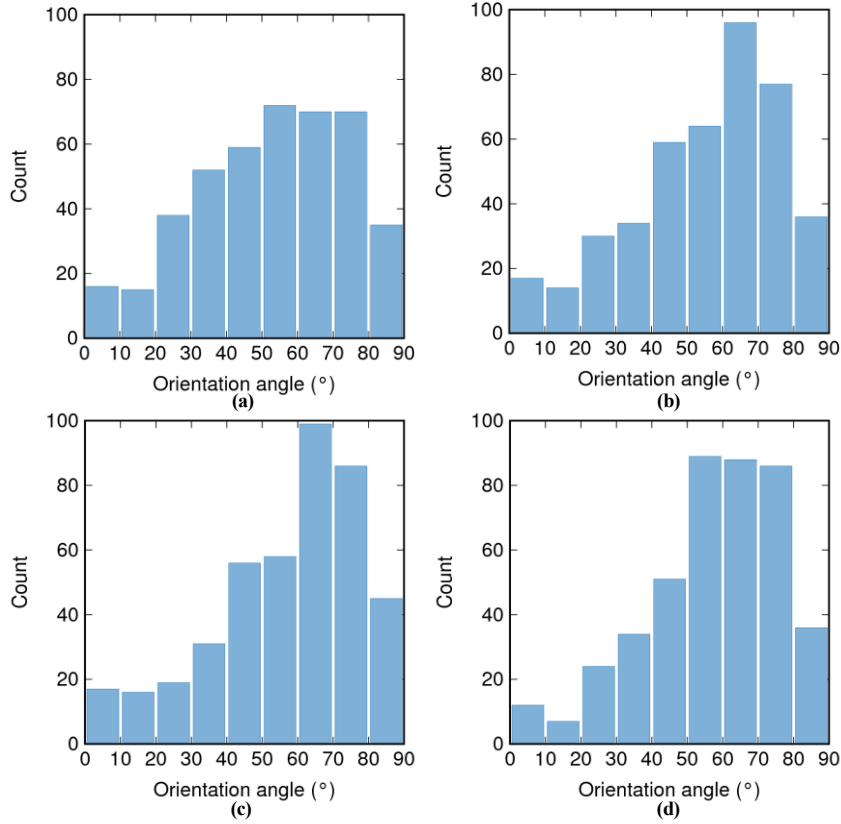

Figure 15: Histogram of the particle orientation angle in the oedometer creep tests under (a) 10 MPa, (b) 100 MPa, (c) 300 MPa, (d) 2000 MPa at the same loading time 0.3 ns.

The "S" shape curve is similar to the laboratory result [57]. The results in Figure16 imply the reliability of our MD simulations of the oedometer creep test at the nanoscale. In what follows, we present the results of the effect of particle sizes on the results of the oedometer creep testing.

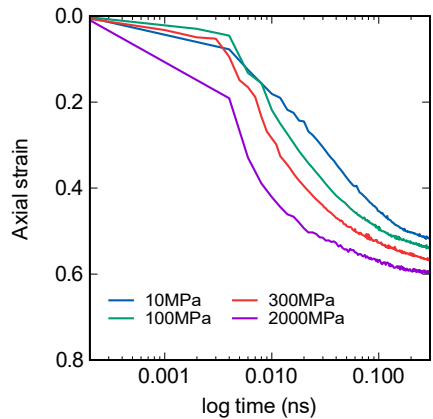

Figure 16: Variation of the axial strain with the logarithmic time during the oedometer creep test.

*3.1.2. Effect of particle sizes*

In this part, we investigate the effect of the hexagonal particle on the nanoscale creep behavior of clay aggregates. Figure 17 presents the initial configurations of three clay samples. The three samples have the same initial dimensions and the same packing density. Sample 1 consists of 427 type-1 particles, sample 2 consists of 889 type-2 particles, and sample 3 consists of 289 type-1 and 289 type-2 particles. As described in Section 2, the type-1 particle is the hexagonal particle with $D = 5.68$ nm, and the type-2 particle refers to the hexagonal particle with $D = 3.87$ nm Figure 18 compares the variation of axial strains for the three samples in the oedometer creep test under 100 MPa. The clay sample with larger particles shows a stronger resistance to creep deformation. For example, at t = 0.3 ns, the axial strains for sample 1, sample 2, and sample 3 are 0.533, 0.570, and 0.551, respectively. We note that the micropores formed by smaller particles can easily collapse through particle rearrangement under the oedometer creep. However, the micropores formed by larger particles involve more face-to-edge configurations of neighboring particles, which are relatively more difficult to collapse under the oedometer creep. To demonstrate this reasoning, Figure 19 compares the deformed structures of the three clay samples at t = 0.3 ns.

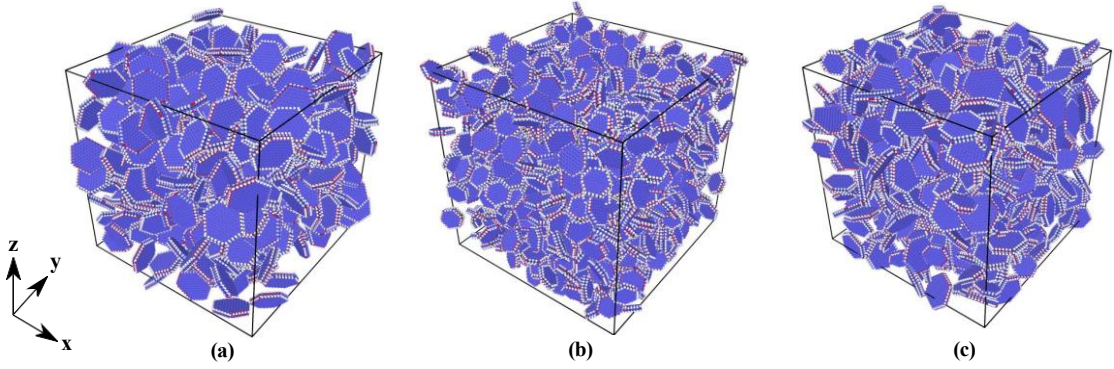

Figure 17: Initial configurations of (a) sample 1, (b) sample 2, and (c) sample 3.

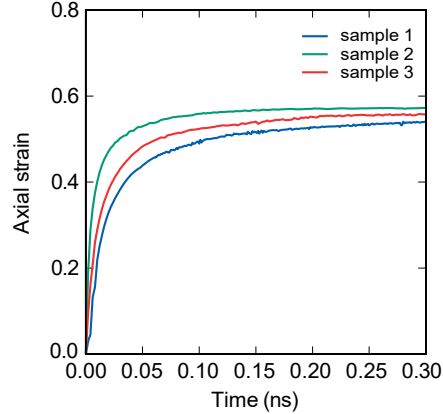

Figure 18: Variation of the axial strain in the three samples.

Figure 20 shows the order parameter *S* variation of the three clay samples in the oedometer creep test under 100 MPa. Sample 1 has the largest order parameter $S = 0.47$, and sample 3 has the smallest order parameter $S = 0.37$. This difference in the order parameter could imply that the clay aggregate consisting of uniform particles favors the more orientated configuration, i.e., the preferred shifted face-to-face structure. However, sample 3 with mixed particles exhibits a more disordered particle arrangement. Figure 21, 22, and 23 present the configurations of selected particles at three loading times, respectively. These five particles are selected at a similar location in the three samples for comparison. Particles in sample 1 exhibit the most ordered arrangement, i.e., shifted face-to-face configurations. By contrast, sample 3 shows more face-to-edge configurations. In the following subsection, we investigate the nanoscale shear creep mechanism by analyzing the micro-structure variation in shear creep testing.

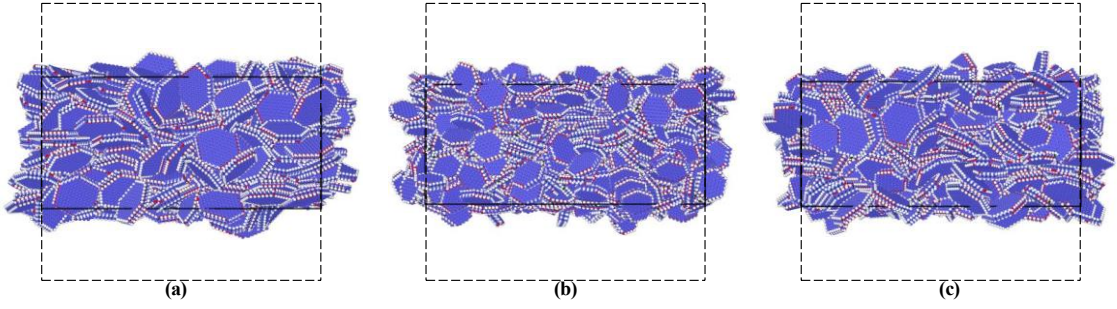

Figure 19: Configurations of (a) sample 1, (b) sample 2, and (c) sample 3 in the oedometer creep test under 100 MPa at t = 0.3 ns.

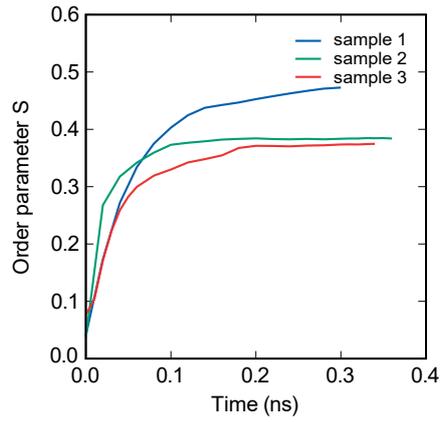

Figure 20: Variation of the order parameter $S$ in the three samples in the oedometer creep test under 100 MPa.

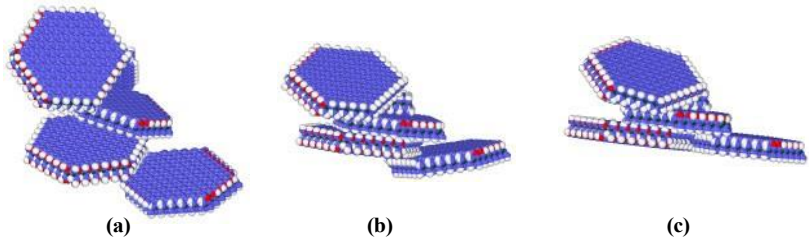

Figure 21: Configurations of selected particles in sample 1 at three loading times: (a) t = 0 ns, (b) t = 0.05 ns, and (c) t = 0.2 ns in the oedometer creep test under 100 MPa.

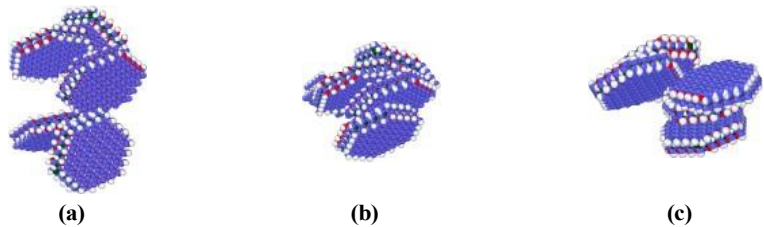

Figure 22: Configurations of selected particles in sample 2 at three loading times: (a) t = 0 ns, (b) t = 0.05 ns, and (c) t = 0.2 ns in the oedometer creep test under 100 MPa.

*3.2. Shear creep test*

This part deals with the results of the shear creep test. Figure 24 presents the variations of the shear strain of clay samples in four shear creep tests under four shear stress levels. The shear stresses are 0.20 GPa in test 1, 0.41 GPa in test 2, 0.813 GPa in test 3, and 4.07 GPa in test 4. In tests 1, 2, and 3, the creep curve shows three typical stages, i.e., primary, secondary, and tertiary. The primary creep corresponds to the initial linear increase part in shear strain. During the second creep stage, the clay aggregate remains in a steady-state creep to a great extent. After the secondary creep, it follows the tertiary creep. A comparison of the shear strain-time curves under the four shear stress levels demonstrates that lower shear stress could accelerate the occurrence of the second creep. For the shear creep test 4 with a higher shear stress of 4.07 GPa, only a steep linear increase of shear strain occurs. This could be due to the microstructure instability caused by the high shear stress.

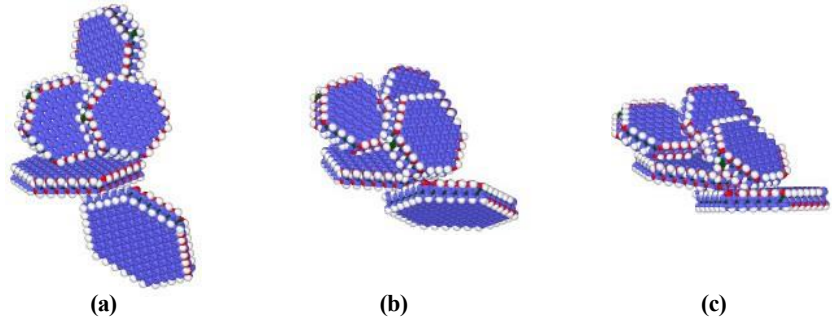

Figure 23: Configurations of selected particles in sample 3 at three loading times: (a) t = 0 ns, (b) t = 0.05 ns, and (c) t = 0.2 ns in the oedometer creep test under 100 MPa.

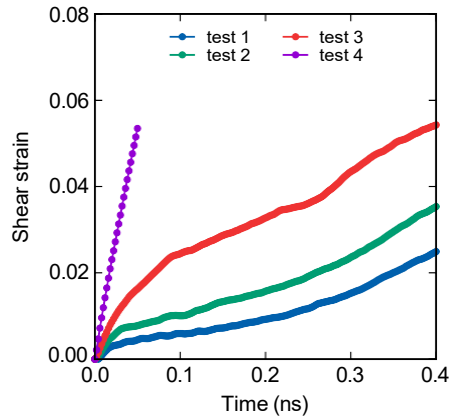

Figure 24: Variation of shear strains in the four shear creep tests.

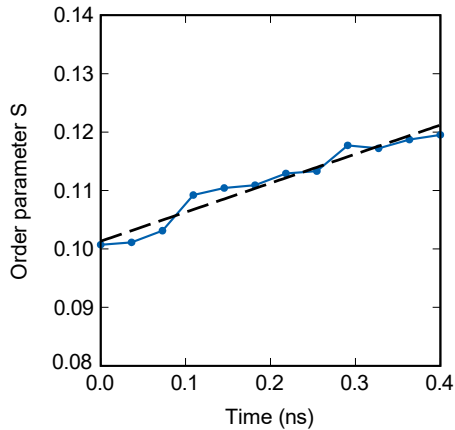

Figure 25: Variations of the order parameter $S$ in the clay sample under shear creep test 3.

In what follows, we investigate particle orientations during the shear creep deformation. For this purpose, in Figure 25, we plot the variation of the order parameter $S$ with time. Since the clay aggregate undergoes very slow deformation, the changing rate of the order parameter is only 0.047 per nanosecond. We use test 3 as an example to characterize the time evolution of particle configurations under shear creep. Figure 26 compares the configurations of four selected clay particles in the shear zone at three loading times t = 0 ns, 0.2 ns, and 0.4 ns during the shear creep test 3. The results show both particle rotation and translation during shear creep. We note that the nanoscale shear creep behaviors in this study are consistent with the rotational and translational slidings observed in the landslide motions (e.g., [9]).

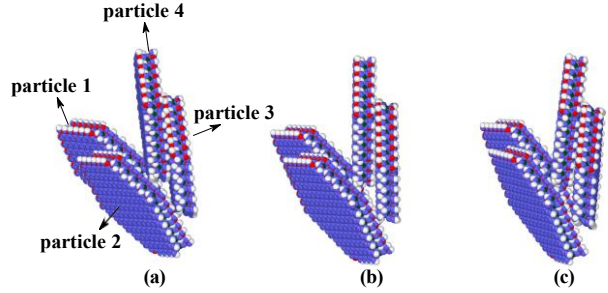

Figure 26: Configurations of selected particles in the shear zone at three loading times: (a) t = 0 ns, (b) t = 0.2 ns, (c) t = 0.4 ns in the shear creep test 3.

In this study, the clay particle rotation in the shear creep test is precisely quantified through the variation of the dihedral angle $\alpha$ between the basal plane of individual clay particles and the x-y plane. Figure 27 shows the schematic illustration of the dihedral angle $\alpha$. Figure 28 plots the variation of the dihedral angle $\alpha$ of selected particles with time. The results show that the average dihedral angle change for selected particles is approximately 10.8° within 0.4 ns.

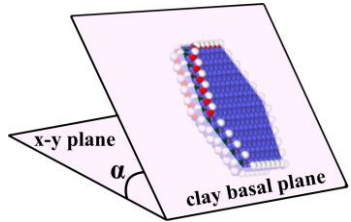

Figure 27: Schematic of the dihedral angle $\alpha$ between the basal plane of a clay aprticle and the x-y plane.

Next, we investigate the loading rate effect on the shear creep behavior of clay aggregates at the nanoscale. We present the results of simulations with two loading rates. For loading rate 1, the shear force is applied immediately. For loading rate 2, the shear stress is applied through a ramp function at 59 kcal/(mol·Å$^3$) per ns. Figure 29 compares the variations of shear strain with time under the two loading rates. The two curves are almost identical in the first 0.25 ns. The results show a noticeable tertiary creep in both tests. However, the test with the incrementally applied shear stress generated less tertiary creep in clay. In the following section, we present the results of the direct shear test of the clay sample.

*3.3. Direct shear test*

In this part, we investigate the effects of shear rates and particle sizes on the shear deformation of the clay aggregate sample under direct shear. In the direct shear test, shear stress is defined as the ratio of horizontal force $F_x$ to the cross-sectional area $A_{xy}$ of the clay sample in the x-y plane. The time evolution of the microstructure of particles in the shear zone is analyzed by characterizing the local particle rotation and the variation of the order parameter of the clay aggregate. We first present the numerical results of the base direct shear test with a shear rate of 0.002 Å /fs. Figure 30 plots the variation of shear stress with respect to shear strain. The results in Figure 30 show that the shear stress increases linearly with the shear strain first and then reaches the peak value around the shear strain of 0.138. After the peak, the shear stress decreases with oscillation under further shear deformation.

Two particles in the shear zone are selected to characterize particle rearrangement under the direct shear test. Figure 31 presents the configurations of two selected particles at different shear strains. Under the global shear loading, the two particles are rotated and translated along the shearing direction. The center-to-center distance between the selected particles decreases, the overlapping area increases, and finally, the two particles form a shifted face-to-face configuration similar to the particle stacking in the oedometer creep test. The dihedral angle $\alpha$ between the two particles is tracked to quantify the magnitude of particle rotations. Figure 32 compares the dihedral $\alpha$ variation of the two particles at the shear rate 0.002 Å /fs. The change of $\alpha$ is up to 60° under the direct shear test. Next, we present the results of the impacts of strain rates and particle sizes on the shear deformation of the clay sample.

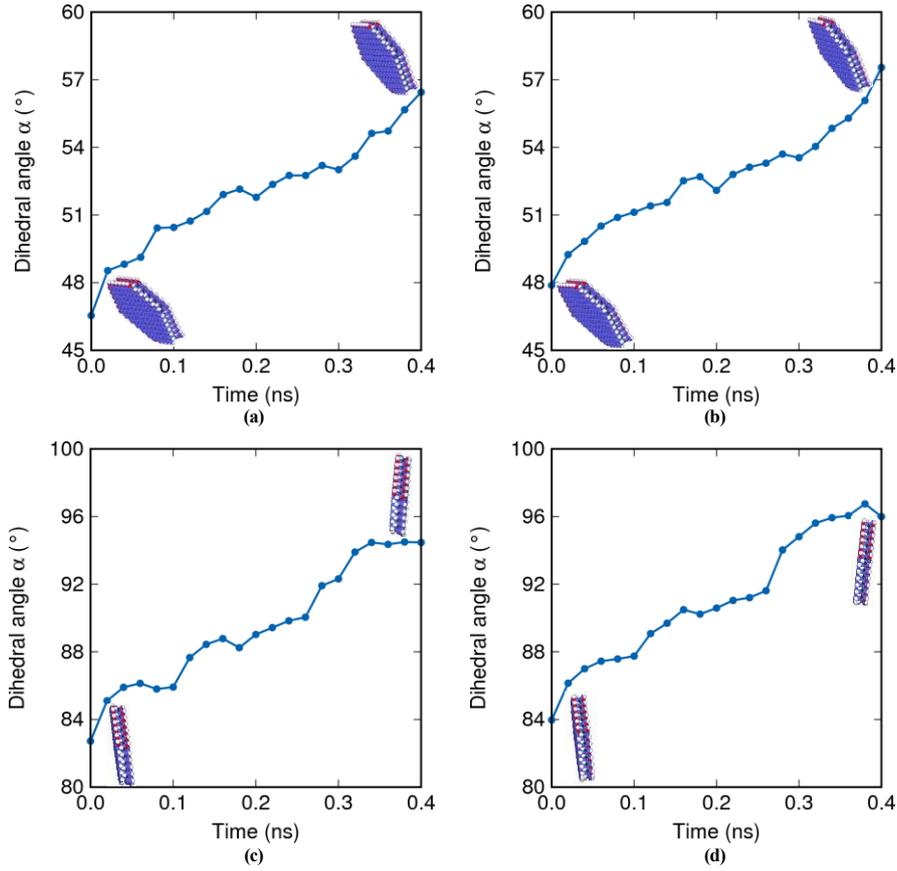

Figure 28: Variations of the dihedral angle with time for (a) particle 1, (b) particle 2, (c) particle 3, and (d) particle 4 in the shear creep test 3.

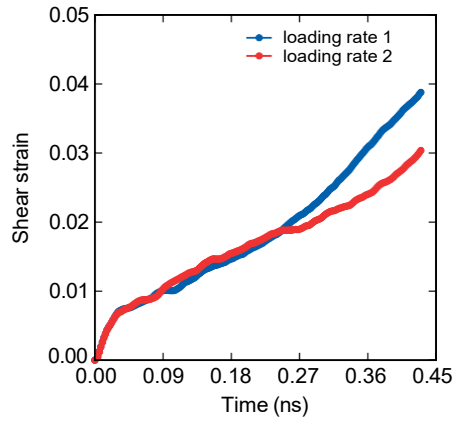

Figure 29: Variations of the shear strain with time under the two loading rates.

*3.3.1. Influence of strain rates*

The direct shear tests under four strain rates are conducted to study the influence of strain rates on the shear deformation of the clay. Figure 33 plots the variation of shear stress with shear strain under four shear rates ranging from 0.001 Å /fs to 0.004 Å /fs. It can be found from the results in Figure 33 the peak stress is dependent on the shear rate, i.e., the higher peak stress with a more significant strain rate. Specifically, the peak shear stresses are 9.87 GPa, 29.06 GPa, 66.10 GPa, and 126.79 GPa at the rates, 0.001 Å /fs, 0.002 Å/fs, 0.003 Å/fs, and 0.004 Å/fs, respectively.

The order parameter *S* is used to quantify particle orientations. Figure 34 shows the variation of *S* with shear strain. It is found that the order parameter first decreases and then increases. This phenomenon is consistent with the observation in Kaolin under the undrained shear test of Kaolin through digital image analysis of both optical and scanning electron micrographs Bai and Smart [63]. Figure 35 compares the variations of the dihedral angle $\alpha$ under four shear strain rates. Despite different shear strain rates, the increase in the dihedral angle of the two particles is similar. It might be implied from the observation that particle rotations are less dependent on the shear strain rates considered in this study.

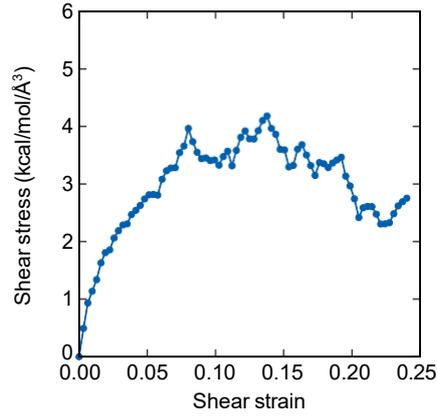

Figure 30: Variation of shear stress with shear strain in the direct shear test under shear rate 0.002 Å/fs.

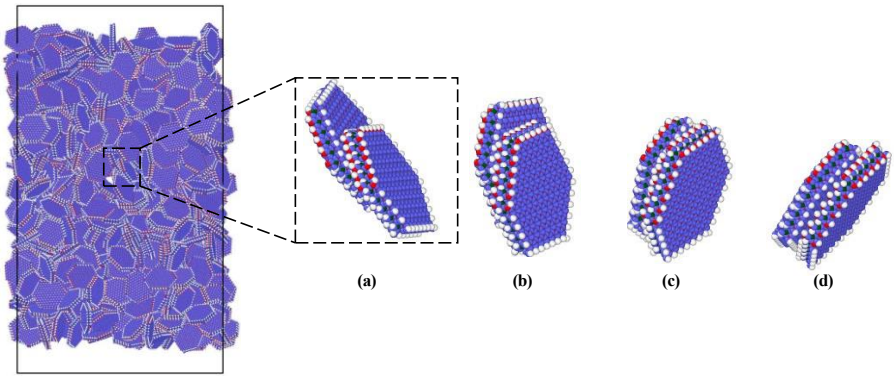

Figure 31: Configurations of two selected particles at four shear strains: (a) 0, (b) 0.077, (c) 0.154, and (d) 0.231 during the direct shear test.

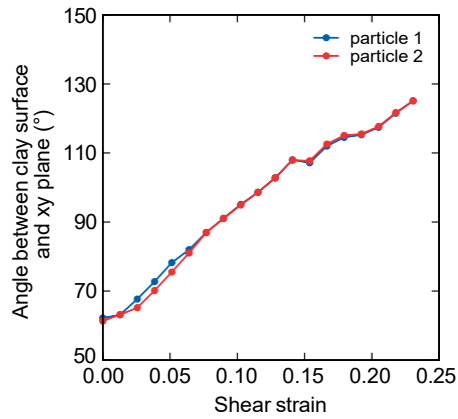

Figure 32: Variations of the dihedral angle $\alpha$ with shear strain at the shear rate 0.002 Å/fs under the direct shear test.

*3.3.2. Influence of particle sizes*

In this part, we analyze the particle size effect on the shear strength and microstructure of clay samples under the direct shear test. As summarized in Table 1, sample 4 consists of 523 type-1 particles, and sample 5 consists of 1088 type-2 particles. Figure 36 compares the configurations of the two clay samples at four shear strains. Figure 37 shows the effect of particle sizes on the shear stress-strain curves at four different shear strain rates. Under the same shear strain rate, the sample with larger particles (i.e., 5.68 nm) has greater shear strength and is more sensitive to the shear rate. For example, the results Figure 37 (a) and (d) show that the peak shear stress of sample 4 under the shear rate 0.004 Å/fs is 12 times larger than that under the shear rate 0.001 Å/fs. In the following section, we present the shear stress relaxation test results.

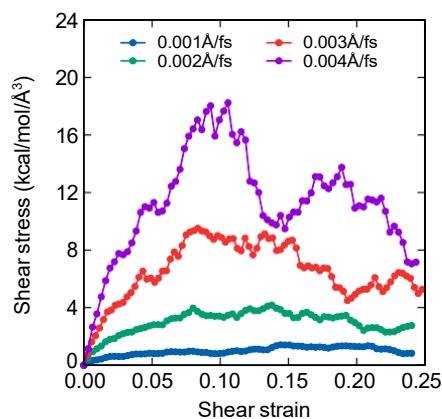

Figure 33: Variations of shear stress with shear strain at four shear strain rates.

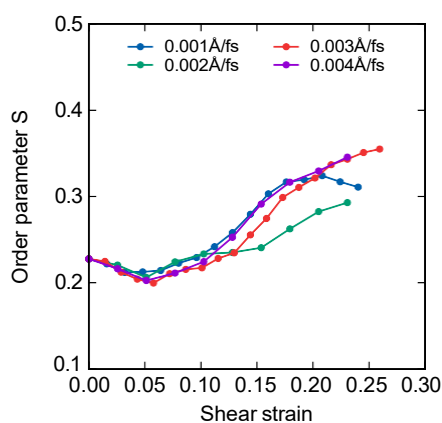

Figure 34: Variation of the order parameter *S* of particles in the shear zone under the direct shear test.

*3.4. Shear stress relaxation*

In this part, we study the shear stress relaxation of the clay aggregate. Stress relaxation refers to the phenomenon that the stress in the sample decreases with time under a fixed strain. The three stress relaxation tests start at different times. Sample 4 is used in the shear stress relaxation tests. We conduct three relaxation tests based on the direct shear test with the shear strain rate 0.002 Å /fs. The three tests correspond to the relaxation at the pre-peak, peak, and post-peak stress under the direct shear test. The specific timings for the relaxation are 0.01 ns, 0.04 ns, and 0.07 ns for test 1, test 2, and test 3, respectively. Figure 38 plots the shear stress variation before and after the stress relaxation. The shear stress of the shear zone particles decreases with time at a declining changing rate. Although the relaxation takes place at different loading times in the three tests, the stress drop and particle movement trends are similar. Figure 39 shows the configurations of five selected particles at different times under the stress relaxation starting at 0.01 ns. Figure 40 shows the displacements of shear zone particles under stress relaxation. In the three tests, shear zone particles translate along the shear direction with increasing horizontal displacement after the termination of shear deformation. After the shear stress relaxation, the vertical displacement continues to increase. After 0.003 ns, we observe the rebound in the vertical displacement of shear zone particles. This delayed rebound may indicate the viscosity of the clay aggregate.

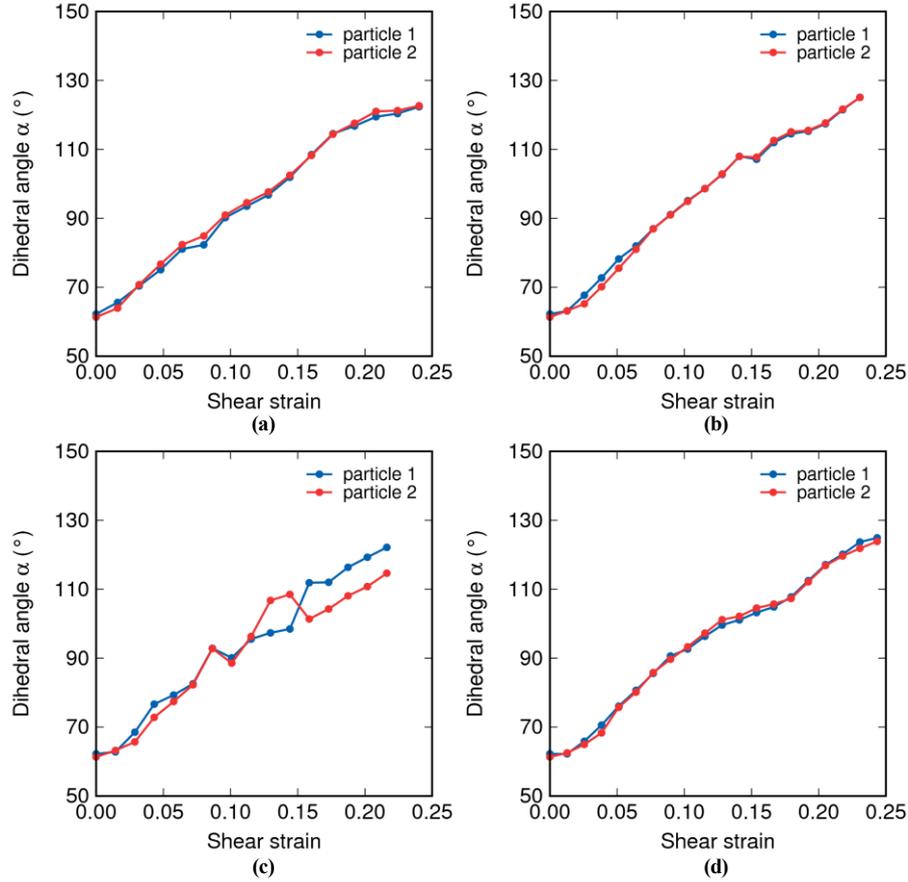

Figure 35: Variations of angle $\alpha$ with shear strain at four shear rates: (a) 0.001 Å/fs, (b) 0.002 Å/fs, (c) 0.003 Å/fs, and (d) 0.004 Å/fs under the direct shear test.

## 4. Concluding remarks

In this study, we have investigated the creep mechanism of clay at the nanoscale. We have conducted MD simulations of the oedometer and shear creep tests and direct shear and relaxation tests of clay samples with two uniform hexagonal particles. The numerical results are analyzed from a microstructure/fabric perspective to discover the nanoscale mechanism of clay creep. It is found from our virtual numerical tests through MD that the creep in clay is due to particle rearrangements under various loading conditions. This study shows that the factors impacting the creep of clay at the nanoscale include particle sizes, stress levels, and loading rates. We note that other factors affecting the nanoscale creep mechanism of clay could be water, chemical solutions, and temperature, which are not the focus of the present study.

For the oedometer creep, the change in the microstructure of clay consists of three stages: (a) the closure of pore space; (b) particle rearrangement, and (c) progressive slippage of interfaces between particles and balanced face-to-edge or face-to-face particle configurations. The transition from process (a) to (b) results from inter-particle interactions as the distance between two particles is within the cutoff distance of pair interaction. Inherently, particle rearrangement is driven by the Lennard-Jones potential and the Coulombic pairwise interaction. With the applied vertical stress increase, the order parameter increases, which indicates a microstructure with more parallel particles.

For the shear creep, the progressive particle rotation and translation occur along the shear direction within the shear zone. The order parameter increases linearly with time at a low rate, implying the progressive formation of the preferred face-to-face configuration. For the direct shear test, the particles near the sliding interface show significant changes in microstructure. Similar to particles in the oedometer test, shifted face-to-face configurations are formed under direct shear. The interface area between two particles increases during the slippage. After two tracked particles obtain a parallel face-to-face structure, they translate and rotate as an entity under shear deformation. Upon the termination of the shear loading, both horizontal and vertical displacements of particles continue in the shear zone. The results imply that the shear stress relaxation phenomenon can be due to a rearrangement of clay particles at the nanoscale.

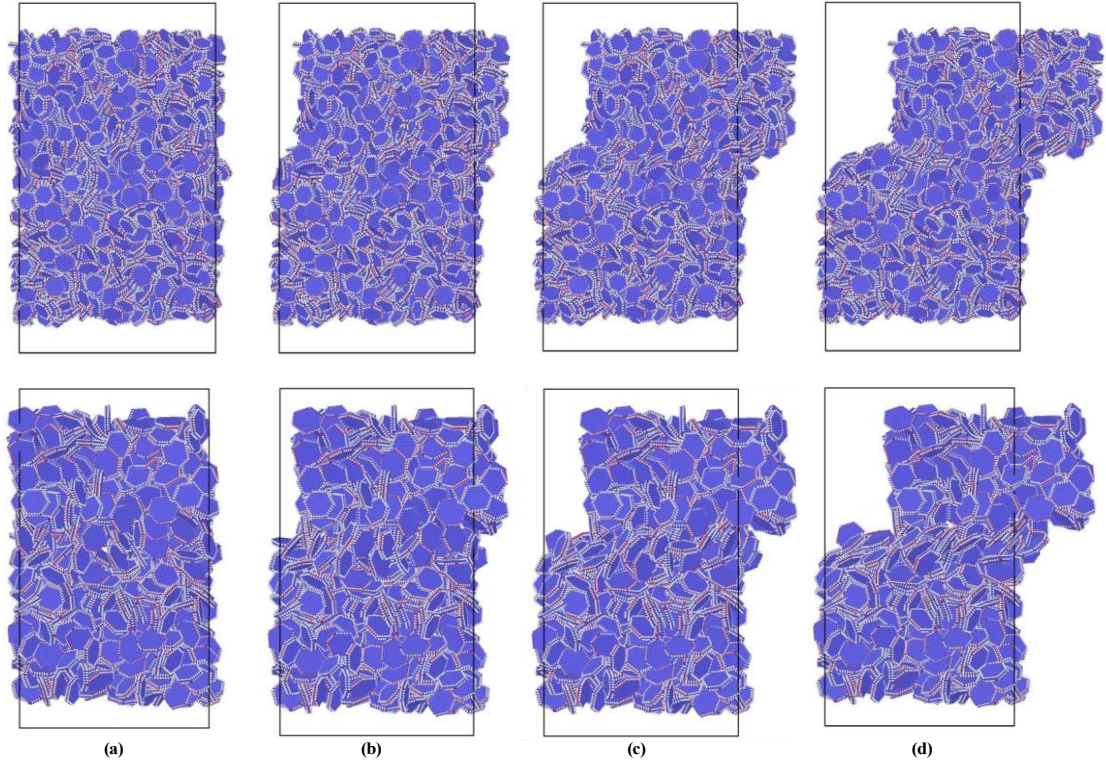

Figure 36: Configurations of sample 1 (the first row) and sample 2 (the second row) at four shear strains: (a) 0, (b) 0.08, (c) 0.16, and (d) 0.24 under the direct shear test.

## Acknowledgment

This work has been partly supported by the US National Science Foundation under contract numbers 1659932 and 1944009. The support is gratefully acknowledged. Any opinions or positions expressed in this article are those of the authors only and do not reflect any opinions or positions of the NSF.

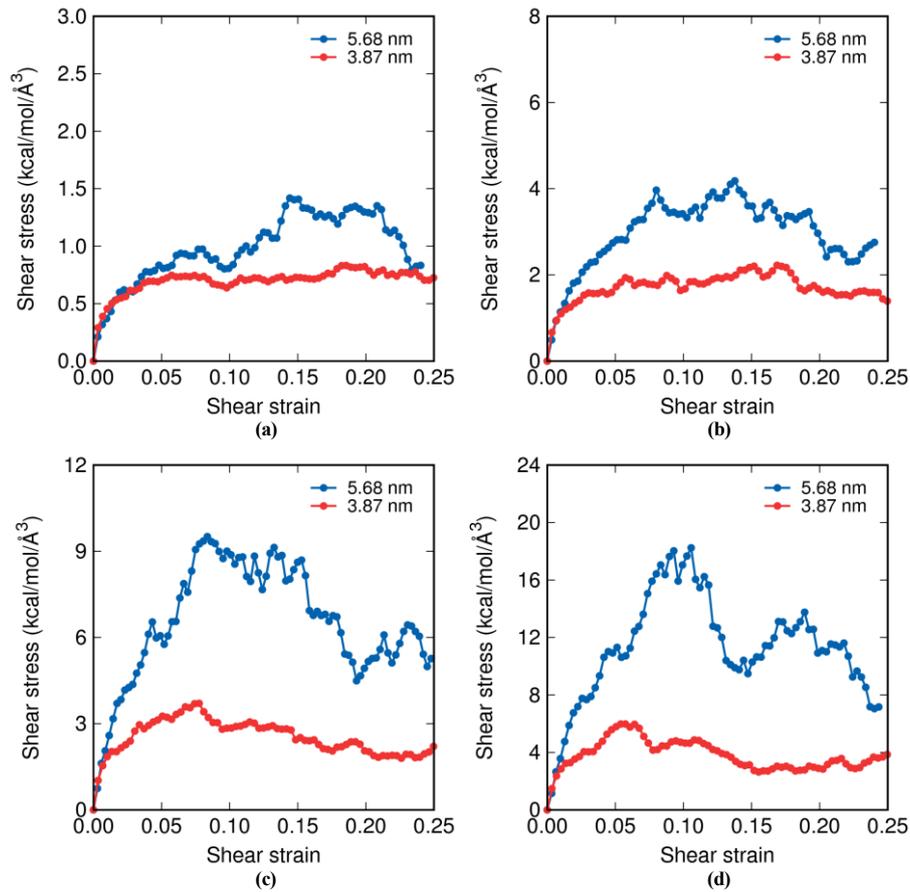

Figure 37: Effect of particle sizes on the shear stress versus strain curve at four shear rates: (a) 0.001 Å/fs, (b) 0.002 Å/fs, (c) 0.003 Å/fs, and (d) 0.004 Å/fs under the direct shear test.

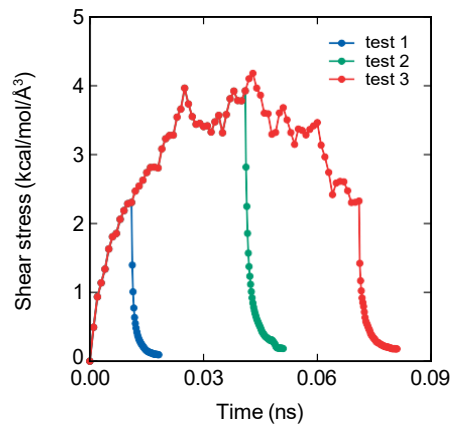

Figure 38: Variations of shear stress in stress relaxation tests.

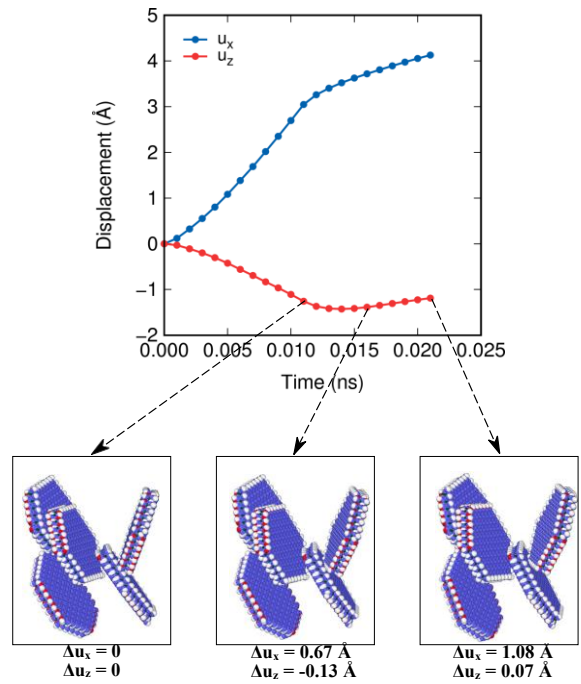

Figure 39: Translation and rotation of clay particles in the shear zone under the stress relaxation.

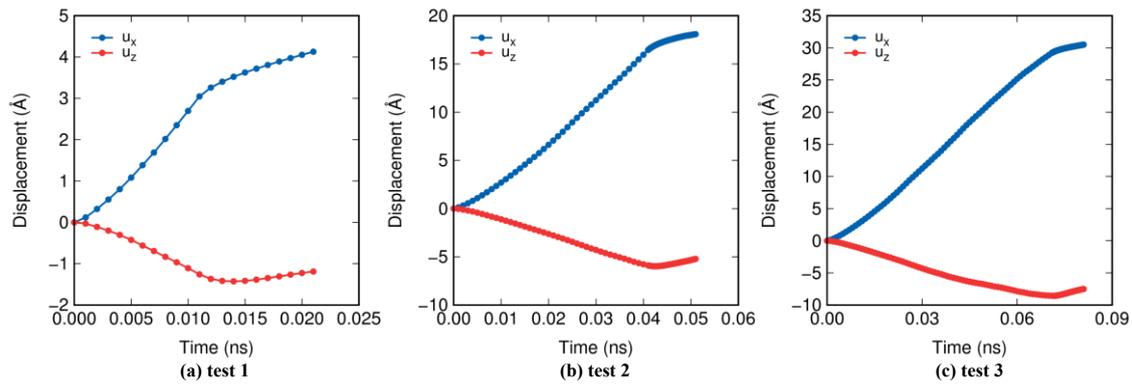

Figure 40: Variations of horizontal and vertical displacements with time of shear zone particles under the stress relaxation tests.